\documentclass[a4paper]{article}

\usepackage[numbers]{natbib}
\usepackage{xurl}
\usepackage{subcaption}
\usepackage{adjustbox}
\usepackage{makecell}
\usepackage{tabularx}
\usepackage{booktabs}
\usepackage{authblk}
\usepackage{multirow}
\usepackage[T1]{fontenc}
\usepackage{lmodern}
\usepackage{floatrow}
\usepackage[a4paper,top=3cm,bottom=2cm,left=3cm,right=3cm,marginparwidth=1.75cm]{geometry}

\DeclareFloatFont{small}{\footnotesize}
\title{The Potential of Sufficiency Measures to Achieve a Fully Renewable Energy System - A case study for Germany}

\author[1,**]{Eerma, M.H.}
\author[1,2,**]{Manning, D.}
\author[1,2,**]{Økland, G.L.}
\author[1,**]{Rodriguez del Angel, C.}
\author[1,2,**]{Seifert, P.E.}
\author[1,**]{Winkler, J.}
\author[1,**]{Zamora Blaumann, A.}
\author[1,*,**]{Zozmann, E.}
\author[1,***]{Hosseinioun, S.}
\author[1,3]{G{\"o}ke, L.}
\author[1,3]{Kendziorski, M.}
\author[1,3]{Von Hirschhausen, C.}

\affil[1]{Technical University of Berlin, Straße des 17. Juni 135, 10623 Berlin, Germany}
\affil[2]{Norwegian University of Science and Technology, NO-7491 Trondheim, Norway}
\affil[3]{German Institute for Economic Research (DIW Berlin), 10117 Berlin, Germany}
\affil[*]{Corresponding author: ez@wip.tu-berlin.de}
\affil[**]{These authors contributed equally to this work}
\affil[***]{These authors contributed to the original research}
\begin{document}

\maketitle

\begin{abstract}
The paper provides energy system-wide estimates of the effects sufficiency measures in different sectors can have on energy supply and system costs. In distinction to energy efficiency, we define sufficiency as behavioral changes to reduce useful energy without significantly reducing utility, for example by adjusting thermostats. By reducing demand, sufficiency measures are a potentially decisive but seldomly considered factor to support the transformation towards a decarbonized energy system.

Therefore, this paper addresses the following question: What is the potential of sufficiency measures and what is their impacts on the supply side of a 100\% renewable energy system? For this purpose, an extensive literature review is conducted to obtain estimates for the effects of different sufficiency measures on final energy demand in Germany. Afterwards, the impact of these measures on the supply side and system costs is quantified using a bottom-up planning model of a renewable energy system. Results indicate that final energy could be reduced by up to 20.5\% and as a result cost reduction between 11.3\% to 25.6\% are conceivable. The greatest potential for sufficiency measures was identified in the heating sector.

\end{abstract}

\section*{Highlights}
\begin{itemize}
	\item Potential of behavioral demand reductions for different sectors is estimated.
	\item The impact of reductions on an exemplary renewable energy system is assessed.
	\item Results suggest that reductions have a significant potential to reduce system costs.
\end{itemize}

\section*{Word count}
\noindent
5673 words (excluding title, author names and affiliations, keywords, abbreviations, table/figure captions, acknowledgements and references) 

\section*{List of abbreviations}
\noindent
CCST \indent Carbon capture, transport and storage \\
TWh \quad \quad Terawatt hours

\section{Introduction} 
Since the Paris Agreement in 2015, annual CO$_2$ emissions have increased by more than 4\% and the remaining emission budget to stay within 1.5\textdegree C of global warming is shrinking rapidly \citep{peters_carbon_2020, mundaca_demand-side_2019}. There is an extensive and growing variety of studies on how to enable a zero-carbon transformation, but the focus of policy advice is on the supply side discussing technological options, like renewable energy, nuclear power, and carbon capture, transport and storage (CCST).

However, according to consumption based accounting, approximately two thirds of global emissions are linked to private households and little effort is made to reduce final energy demand \citep{grubler2018low, SAMADI2017126, emissions2020}. Despite this high potential for climate change mitigation, demand side solutions are still poorly understood.  
Previous work has addressed the importance of demand-side solutions to stay within 1.5\textdegree C, but finds that these are still insufficiently covered by quantitative energy system models \citep{mundaca_demand-side_2019, grubler2018low}. Against this background, this study evaluates the impact of sufficiency measures in an energy system based on 100\% renewable energy.

The term energy sufficiency often refers to an absolute reduction of energy demand. In a more general definition, sufficiency is also understood as a concept that ensures an equitable energy access while staying within planetary boundaries. \citet{darby2018energy} use the term energy sufficiency to refer to 'a state in which humanity would only consume energy services equitably and in quantities compatible with sustainability and ecological limits'. To facilitate a better understanding, sufficiency is commonly differentiated from efficiency and consistency. While efficiency improves the input-output ratio and consistency changes production and consumption patterns (e.g. replacing fossil generation with renewable generation), sufficiency limits absolute energy consumption while still fulfilling basic human needs \citep{darby2018energy, SAMADI2017126}. Throughout this paper, sufficiency measures are henceforth understood as behavioral changes that lead to \textit{reductions of energy consumption to a level where utility is not significantly diminished}, for example lowering thermostats by 1\textdegree C.  

Understanding sufficiency policies and quantifying their effects on the energy system is an important strategy for a zero-carbon transformation because it identifies the potential of behavioral changes and increases the likelihood of their realization. Emphasized by experts and scientists in the field of sufficiency, \citet{Toulouse2019} indicate that incorporating sufficiency in energy system modelling and defining quantitative scenarios for it is necessary to provide objective criteria for policy-makers. Furthermore, reviewing existing prominent energy scenarios illustrates that most do not include sufficiency, concluding that it should be more evaluated and incorporated in energy system modelling
\citep{SAMADI2017126,Pfenninger2014}.

There are several studies modeling a decarbonized Germany energy system. While some studies incorporate lifestyle changes in one of their scenarios, a majority of studies do not consider any lifestyle changes at all \citep{Fraunhofer2020, greiner2016sektorale, uba_2019_rescue, uba_2018_sufficiency_models}. This paper seeks to fill this gap by explicitly defining the potential of sufficiency-based demand reductions and their impacts on the supply side of a 100\% renewable energy system in Germany. Based on an extensive literature review, the potential of sufficiency measures for the heat, mobility and electricity sector is estimated. A cost minimizing bottom-up planning model is applied to estimate the impacts of these reductions on a greenfield, renewable energy supply for Germany in a scenario-based approach. The model is implemented in the AnyMOD Framework \citep{goke2020anymod, goke2020anymodjl} with renewable availability time series in an hourly temporal resolution and technology cost assumptions for 2035. 

This remainder of this paper is organized into five sections. The next section identifies the potential for sufficiency measures for the heat, mobility, and electricity sector. Section 3 lays out the energy system model applied and introduces two scenarios with different levels of ambition regarding sufficiency. Section 4 presents and discusses the results, Section 5 concludes.

\section{Quantifying the potential of sufficiency measures}\label{sec:theoretical_background}
\begin{figure}
    \centering
    \includegraphics[width=0.7\textwidth]{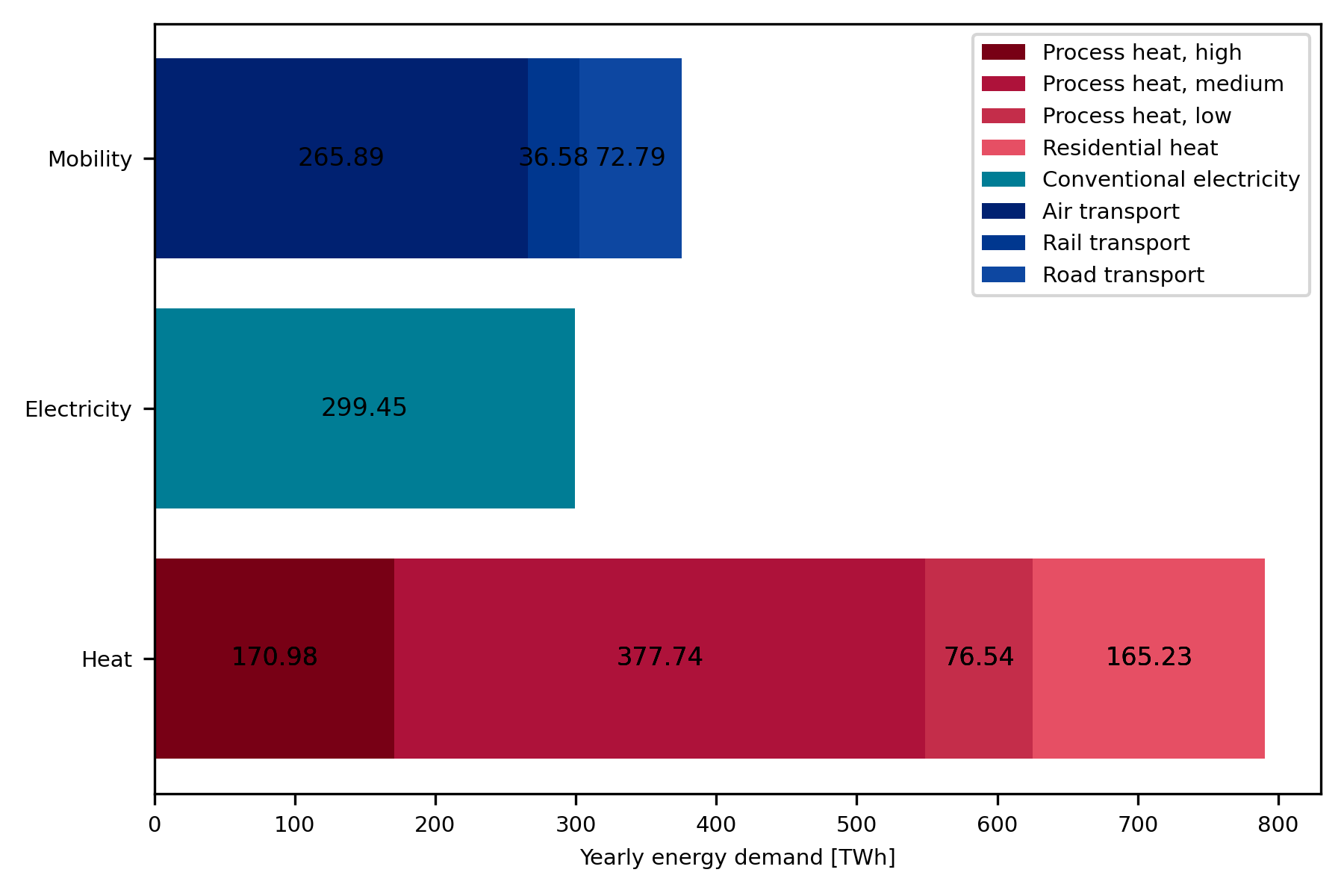}
    \caption{Annual energy demand 2050 in terawatt hours [TWh] by sector. Color shade and black numbers represent the subcategories of the respective data. Own illustration based on \citet{auer2020development}.}
    \label{fig:demand_sub}
\end{figure}

The potential for sufficiency-based demand for Germany in the the heat, mobility, and electricity sector is quantified based on existing literature. Each of the following subsections provides an overview on demand reduction potentials for one sector and concludes with an overall potential for the sector. Most of the demand reductions are only valid for a specific subset of energy demand, e.g. assuming a reduction in short-distance passenger flights only reduces the energy demand of private aviation. Figure \ref{fig:demand_sub} gives an overview of the reference energy demand data used in this paper. The data is taken from the \textit{openENTRANCE} project that develops scenarios compliant with the Paris Agreement \citep{auer2020development}. The scenario 'societal commitment' was chosen, which aims for a strong change of behavior towards a sustainable lifestyle. This scenario already includes energy demand reductions in transport and delivering services, combined with advantages of intense digitalization and first steps towards a circular economy. Given that the demand data is for 2050 and that it already assumes a certain bias towards a more sustainable lifestyle might influence the model results. However, this data was selected since it is one of the few open-source data sets that include time series for all sectors in a high temporal resolution. Furthermore, the focus of this paper lies on the impacts of sufficiency-based demand reductions, hence this drawback was considered to be negligible. Demand data from this project has also been cross-checked for consistency with other publications, most notably the SET-NAV project \citep{auer2020development}. Demand data is differentiated in the sectors mobility, heat and electricity, which are further divided into subcategories (as shown in Figure \ref{fig:demand_sub}). The mobility sector uses electricity for road and rail transport and hydrogen for aviation. The heat sector includes electricity for residential heat and parts of process heat, synthetic gas in process heat, and hydrogen in low-temperature process heat. The potential for sufficiency measures is quantified based on this data, meaning that the overall potentials at the end of each subsection are derived by linking the reductions from literature to the respective shares of energy demand. To account for variations in the scale of demand reductions, the potential also varies by ambition of behavioral changes, hereafter referred to as the \textit{Low Ambition} and \textit{High Ambition} scenario.

\subsection{Heat sector}
The heat sector accounts for 54\% of overall energy demand, reflecting its high importance for sufficiency measures. The sector is subdivided into residential-commercial heating and process heat demand, with the latter subdivided by temperature levels. The conceivable sufficiency-based measures are  are summarized in Table \ref{tab:entire_heat_literature} in the Appendix. For the \textit{Low Ambition} and \textit{High Ambition} scenario the reduction potential is depicted in Table \ref{tab:saving_potential_heat}. Negative percentages refer to the reductions that are derived for the respective category based on literature, differentiated for the \textit{Low Ambition} and \textit{High Ambition} scenario. Percentages in the column headings refer to the share of total heat demand, e.g. residential/commercial heat energy demand accounts for approximately 21 percent of total heat energy demand.

\begin{table}
\caption{Low and High Ambition reduction potential for residential/commercial and process heat energy demand.}
\label{tab:saving_potential_heat}
\resizebox{\textwidth}{!}{
\begin{tabular}{ccccccc}
\toprule
\textbf{Reduction potential} & \textbf{\begin{tabular}[c]{@{}c@{}}Residential/ \\ commercial (20.9\%)\end{tabular}} & \textbf{\begin{tabular}[c]{@{}c@{}}Process heat\\low (9.7\%)\end{tabular}} & \textbf{\begin{tabular}[c]{@{}c@{}}Process heat\\mid (47.8\%)\end{tabular}} & \textbf{\begin{tabular}[c]{@{}c@{}}Process heat\\high (21.6\%)\end{tabular}} &\textbf{Total Heat}\\\midrule
Low ambition &  -5.0\%  & -8.6\% & -5.1\%  & -3.0\% &5.0\%\\
High ambition &  -34.2\% & -13.2\%  & -12.0\%  & -7.6\% &15.8\% \\
\bottomrule
\end{tabular}
}
\end{table}

\subsubsection{Residential and commercial heat}\label{sec:heat_literature}
The average room temperature for Germany's households is 19.6\textdegree C \citep{statista_raumtemperatur}. It is assumed that a reduction of 1\textdegree C is acceptable without diminishing the living comfort. The German Umweltbundesamt finds a potential of 4.4\% to 9\% in 2030 by reducing the temperature by 1 or 2 \textdegree C, taking into account the refurbishment rate in Germany \citep{umweltbundesamt_konzept_2015}. A report from Cambridge Architectural Research estimates the energy saving potential of small changes in household behavior based on a building simulation. They identify a reduction potential of 13\% of the space heating energy by turning down the thermostat by 1 \textdegree C \citep{palmer2012much}.  \citet{marshall_combining_2016} calculate energy savings considering four different measures for three different occupancy patterns. Energy savings of 9\% are achievable by reducing the households' space temperature by 1\textdegree C.

\citet{palmer2012much} argue that the installation of water-efficient shower heads can save up to half of the energy usage for showering. An additional change in behavior caused by climate awareness or price incentives can reduce energy demand for hot water by approximately 20-30\%, as a consequence of shorter and less frequent showering. \citet{toulouse_2018_products} present an average cut of 5-10\% in energy usage, which emerges from feedback about showering time. As stated by \citet{lehmann2015stromeinspareffekte}, the demand for hot water in households can be reduced through shortening the daily shower time and adjusting the water consumption with water-saving fittings by up to 70\%.

\citet{BierwirthThomas2019, bierwirth_energy_nodate} argue that the European trend of increasing living space per person creates high energy demand. Their analysis focuses on the balance between a sufficient floor space for a decent living. The average living space 2018 in Germany is 46.7m$^{2}$ per person \citep{destatis_floorspace}.
Taking into account five different sizes of households, \citet{BierwirthThomas2019} calculate an 'adequate' average space of 32.3m$^{2}$ per person. Consequently, an energy saving potential of 24.9\% (35m$^{2}$/cap) to 35.7\% (30m$^{2}$/cap) is assumed to be reasonable for space heating. Associated reductions of energy service demand related to lighting and building material are not included and would lead to even more saving potentials. Compared to other European countries, energy sufficiency potential of Germany is rated as very high and is only surpassed by Luxembourg \citep{BierwirthThomas2019}.

\subsubsection{Process heat}\label{ph}
Sufficiency measures in the industry sector are for example when energy intensive materials are substituted by less energy intensive products \citep{vita2019environmental}. Energy demand for process heat is subdivided by industry branch according to \citet{auer2020development}, which serves as the baseline demand shown in Table \ref{tab:industrybranches}. It is assumed that each of the following sufficiency measures only affects its respective industry sector. Approximately 12\% of total process heat demand relate to low-temperature heat; 60\% to mid-temperature, and 27\% to high-temperature. The following discusses sufficiency measures for process heat, with one paragraph per temperature level.

\begin{table}[hbt!]
\caption{Process heat demand share of industry branches. Process heat is differentiated into low, mid and high temperature. Percentages refer to the shares of the industry branch on each temperature level \citep{auer2020development}}
\label{tab:industrybranches}

\begin{adjustbox}{max width=\textwidth}
\begin{tabular}{lccccccccc}
\toprule
   \textbf{Temperature}   &\textbf{ Food }   & \textbf{Paper}   &  \textbf{Chemical} & \begin{tabular}[c]{@{}c@{}} \textbf{Engineering /} \\  \textbf{Manufacturing} \end{tabular} & \textbf{Refineries} & \textbf{Other}   & \begin{tabular}[c]{@{}c@{}} \textbf{Non-metallic / }\\ \textbf{ Minerals} \end{tabular} & \textbf{Iron / Steel} & \begin{tabular}[c]{@{}c@{}} \textbf{Non-ferrous /} \\ \textbf{ Metals} \end{tabular} \\ \midrule
Low < 100\textdegree C  & 49.0\% & 51.0\% & 0  & 0   & 0    & 0 & 0   & 0& 0    \\ 
Mid < 500 \textdegree C  & 0 & 0 & 41.3\%  & 20.6\%   & 12.4\%    & 12.3\% & 13.41\%   & 0 & 0     \\ 
High > 500 \textdegree C  & 0 & 0 & 0  & 0   & 0    & 0 & 0   & 85.9\%& 14.1\%  \\ \bottomrule
\end{tabular}
\end{adjustbox}
\end{table} 

As shown in Table \ref{tab:industrybranches}, the food industry accounts for almost 50\% of the low-temperature demand. The current level of food waste could be reduced from 35\% to 17.5\% to reduce the energy demand of the food industry \citep{lebensmittel2019}. A more ambitious assumption of limiting food consumption to 2586 kcal/day leads to a energy demand reduction of 27\% \citep{vita2019environmental}. These sufficiency measures translate to 8.6\% and 13.2\% reduction potential of low-temperature heat demand in \textit{Low Ambition} and \textit{High Ambition}, respectively. 

While the recycling rate for plastics currently only amounts to 47\%, the rate for paper and metal products is 75.9\% and 92\%, respectively \citep{uba_kunststoffe, chemiewirtschaft}. Assuming that plastic recycling can reach similar recycling rates as paper and metal, mid-temperature heat demand can be reduced by 1.4\% in \textit{Low Ambition} and 2.1\% in \textit{High Ambition}. Another key sufficiency strategy mentioned in many studies is an extended product lifetime \citep{umweltbundesamt_konzept_2015, wieser2016beyond}. Implementing a statutory warranty of 5 years and setting the defects liability to 10 years could save at least 14.8\% of the produced products. Additionally, regulating a mandatory availability of components for at least 20 years would ensure prospective repairability. Assuming a similar substitution quota for all product categories reduces the manufacturing/engineering energy demand by 14.8\%, leading to an overall mid-temperature heat demand reduction of 3\% \citep{obsolescenceUBA}. Another potential to decrease energy demand through changed consumption patterns is by supporting consumption communities as an example of shared economies. Research carried out by \citet{vita2019environmental} finds a consumption reduction of 50\%, where 10\% will be shifted to services. Accordingly, the remaining 40\% can be attributed to sufficiency measures, resulting in a sufficiency potential of 8.2\% in mid-temperature heat.

Promoting a modal shift in the construction industry from steel and cement to wooden structures could significantly reduce industrial energy demand, as the iron and steel production makes up 86\% of high-temperature heat demand. According to \citet{hertwich2019material}, 10\% of all construction materials can be replaced by wood already. By assuming such replacement rates for steel in the construction industry, a sufficiency potential of 3\% could be achieved. Based on a reduced average floor space per capita in the residential and commercial heating, as discussed in \ref{sec:heat_literature}, a similar reduction in construction materials is assumed. This leads to a potential demand reduction of 4.6\% in high-temperature heat demand due to reduced steel production and a potential demand reduction of 1\% in mid-temperature heat demand due to decreased cement production. For a more detailed description of the calculation process for process heat, see \citet{pjs_report}.

\subsection{Transport sector}
\label{sec:Mobility}
Despite increases in energy efficiency in the mobility sector, its energy consumption has been rising continuously since 2010 \citep{bmwi_2019_energieeffizienz}. The German government committed to a reduction of final energy consumption in the mobility sector of 15\% to 20\% compared to 2005 levels to achieve reductions of 60\% by 2050 in total energy demand. Mobility demand is divided in rail (19.4\%), road (9.8\%), and air (70.8\%) transport, which is further separated in passenger and freight transport. While rail and road technologies are expected to run with electricity only, air transport demand is fully met by hydrogen \citep{auer2020development}.   

While there is a variety of narratives of a future mobility system, little research explicitly investigates absolute demand reductions. \citet{umweltbundesamt_konzept_2015} discusses four measures with regards to the transport sector: A modal shift to bicycles, replacing business trips with telemeetings, smaller passenger cars and a reduction of private aviation. \citet{van2018potential} model the potential impacts of behavioral changes on climate change mitigation, including several measures in the transport sector. A demand reduction of 30 and 55\% in 2050 for motorized individual transport and aviation respectively is modeled by \citet{Fraunhofer2020}. Two other studies examine the socio-technical transformation of the Danish transport sector, including changes induced by different human behavior \citep{VENTURINIa, VENTURINIb}. Table \ref{tab:mobility} in the Appendix provides an overview of these measures. In the following, sufficiency measures for road, rail and aviation transport are discussed.

\subsubsection{Road}
Passenger cars are the strongest driver of energy demand in the non-commercial mobility sector. If shorter distances are covered with bicycles, hereafter referred to as a modal shift, the energy demand for cars could be reduced by up to 10\% \citep{umweltbundesamt_konzept_2015, van2018potential}. According to the statistics published by the German Ministry of Transport, 18\% of passenger car transport is due to commuting to work \citep{BMVI_verkehr_2019}. Assuming that only one day of home office is possible each week, transport demand related to commuting could be reduced by 20\% \citep{van2018potential}. A further 19\% of passenger car transport is due to business travels \citep{BMVI_verkehr_2019}. Consequently, transport demand related to business travels could be reduced by 60\%, assuming that business trips are gradually replaced by telemeetings \citep{umweltbundesamt_konzept_2015}. Energy demand of passenger car transport is also influenced by the size of private vehicles. Regulations or a changed consumer preference towards smaller private vehicles could therefore potentially reduce this demand by 7.5\%. Finally, more energy-efficient driving patterns could reduce the energy demand of cars by 5\% \citep{van2018potential}.

The German government aims to reduce energy consumption of the total freight transport in Germany by up to 20\% in 2030, which includes both road, rail and air freight transport \citep{bundesregierung_nachhaltigkeitsstrategie_2016}. Training on efficient driving can lead to reduced energy consumption of 10\% \citep{villalobos2016freight}. Online shopping contributes to a large proportion of freight transport by road, as delivery vans around the country are vital for the last mile distribution of shopped goods. Better quality checks for defective items and responsible online shopping has the potential to reduce the rate of return from online shopping by 50\% \citep{pjs_report}.

\subsubsection{Rail}
The identified sufficiency measures for passenger rail transport follow a similar pattern as private car reductions. A modal shift from public transport to bicycles could reduce the energy demand of rail transport by roughly 3\% \citep{greiner2016sektorale}. 21\% of public transport is due to commuting and 11\% of public transport is due to business travels \citep{BMVI_verkehr_2019}. Analogous to passenger car transport, these shares of public transport demand are assumed to be reduced by 20\% through home office and 60\% through telemeetings respectively.

Rail freight transport has seen low levels of absolute growth in the EU compared to the road freight transport since 2000 \citep{dionori2015freight}. The German government has identified the potential of trains in moving freight to reduce CO\textsubscript{2} emissions, as it emits the lowest amount of greenhouse gas emissions per ton-km \citep{lena}. Train speed control during operation has one of the highest potentials for reducing energy use. Controlling parameters, such as acceleration, cruising, and coasting, through optimization algorithms is capable of decreasing the present energy consumption by rail between 20-30\% \citep{corlu2020optimizing}. Using network optimization algorithms can also contribute to reducing energy consumption in rail freight transport. This would require support systems from the government, such as increased public investment in rail infrastructure and expansion planning based on time table intervals coordination \citep{lena}. Such optimization algorithms have the potential to reduce energy demand in rail freight scheduling by up to 14\% \citep{corlu2020optimizing}.

\subsubsection{Aviation}
Aviation constitutes the greatest share of the mobility energy demand, suggesting that behavioral changes will show the most significant impact. In 2020, business flights decreased by 87\% and private flights by 74\%, most likely due to Covid-19 \citep{eurostat, fraport_2020}. The pandemic has indicated that a lot of business meetings can actually be replaced by telemeetings. Furthermore, the well-developed infrastructure and the technological progress in today’s train traffic allow the replacement of national business flights by rail. Up to 60\% of passenger flights in Germany will be related to business travel in 2030 \citep{umweltbundesamt_konzept_2015}. Assuming that the share of passenger flights related to business travels will gradually be replaced by telemeetings to the level that is possible already today, a significant saving potential is identified. Furthermore, assuming that flight behavior in Germany will change towards more sustainable behavior, the private flight rate of 2020 was taken as a basis for the reduction potentials. A total decline of 54\% in passenger air demand is calculated.

Increasing freight traffic decreases the likelihood of significant energy reductions in freight aviation demand. A ton of freight transported by train instead of plane currently consumes more than 90\% less energy and the energy consumption per ton of freight for ships is even lower.\footnote{Own calculations. For a detailed description of the calculation process, see \citet{pjs_report}.} In 2019, 22\% of all cargo flights to and from Germany were continental flights (EU-27) \citep{cargo_2019}. Assuming that these flights could be replaced by ships and trains constitutes a saving potential of 20\% in air freight transport in \textit{Low Ambition} and approximately 30\% in \textit{High Ambition}, assuming that 10\% of intercontinental flights could also be replaced. 

The aggregated \textit{Low Ambition} and \textit{High Ambition} potential for the mobility sector is summarized in Table \ref{tab:saving_potential_mobility}. It is derived by linking the discussed reductions with the respective shares of energy demand. Again, a negative percentage refers to the reductions that are derived for the respective category and percentages in the column headings refer to the share of total mobility demand, e.g. rail mobility energy demand accounts for approximately 10 percent of total mobility energy demand.

\begin{table}
\caption{Low and High Ambition reduction potential for air, road and rail mobility energy demand.}
\label{tab:saving_potential_mobility}
\centering
\small
\begin{tabular}{@{}ccccc@{}}
\toprule
\textbf{Reduction potential} & \textbf{Air (71\%)} & \textbf{Road (19\%)} & \textbf{Rail (10\%)} & \textbf{Total Mobility} \\ \midrule
Low Ambition      & -21.6\%              & -17.7\%               & -30.2\%               & -21.7\%                  \\
High Ambition      & -31.9\%              & -21.3\%               & -41.9\%               & -30.9\%                  \\ \bottomrule
\end{tabular}
\end{table}

\subsection{Electricity sector \label{sec:convetional}}
The sector includes electricity used for traditional applications such as lighting, brown goods like information technology, white goods like refrigeration, and mechanical energy. So far, efficiency has played an important role in reducing conventional electricity consumption. However, growing proliferation of appliances, increase in sizes and functionalities, longer usage hours, and new areas of application can be observed \citep{toulouse_2018_products}. This can be partially attributed to rebound effects because when appliances become more efficient and therefore less energy-consuming, households can spend the saved income on other energy-intensive goods or increase the usage time of existing commodities \citep{toulouse_2018_products, sorrell2020}. This perspective indicates that efficiency alone will not be enough to achieve climate protection goals and societal behavior will play a decisive role the transformation of the energy system \citep{Fraunhofer2020}.

The total conventional electricity consumption is distributed between households (25.3\%), commercial spaces (28.9\%) and industry (45.8\%) \citep{bmwi_2019_energieeffizienz}. The highest share of electricity in households is used for low process heat (i.e. cooking and baking) and cooling. Also, information technology has a considerable share in electricity consumption. In the commercial sector, most electricity is used for lighting and other appliances, such as computers. In industry, the use of electricity to operate electric-driven motors and machines dominates the demand. The partition of households, commercial, and industry is used to identify measures in the literature and quantify their respective potential for demand reductions. 

The literature focusing on sufficiency in the electricity sector is scarce. While most of the studies assess the impact of efficient technology on demand, few evaluate the total potential of sufficiency actions regarding conventional electricity. As stated in \citet{umweltbundesamt_konzept_2015}, the main problem that arises in the evaluation of electricity reduction potentials is the partial overlapping of some implemented measures, making it difficult to sum them accurately. Table \ref{tab:electricity} in the Appendix presents an overview of the instruments and measures applied and their resulting reduction potential that were identified in the literature. Sufficiency measures are differentiated by the instruments to incite behavioral changes. For households, feedback systems seem to have a considerable effect in motivating households to reduce their electricity consumption.  \citet{martiskainen_affecting_2007} use direct feedback, either from a smart meter or a display monitor, to assess the impact of knowledge on the amount of electricity consumed. They state that up to 15\% of electricity can be saved together with setting a reduction goal. A more recent study by \citet{en12193788} estimates an average reduction potential by examining 64 studies mainly in Europe and North-America on different feedback applications, e.g. through in-home displays, load monitors, smart hubs, or energy portals, and state either a 9\% reduction potential through direct or 4\% through indirect feedback. \citet{burger2009identifikation} has conducted one of the few country-level studies that evaluates the maximum achievable decrease in electricity use by "lining up" various sufficiency measures, such as optimizing the usage of different devices and avoiding stand-by losses. According to \citet{burger2009identifikation}, 20\% of electricity could be saved in Germany. A similar approach is used in a study by \citet{umweltbundesamt_konzept_2015}, which summarizes the energy demand reduction potential by 2030 in several sectors in Germany, including household appliances.

Energy consumption from non-residential buildings, like commercial offices, is more difficult to assess compared to typical households due to different building sizes, varieties of activities and since employees mostly share common equipment, making them feel less responsible for conserving energy \citep{CARRICO20111}. For commercial spaces, \citet{CARRICO20111} and \citet{nilsson2015energy} both use group feedback systems and peer education as tools to change employee behavior and find 7\% and 6\% reduction in electricity consumption. Results in \citet{hansen_working4utah_2009}  show an overall energy demand reduction of 10.5\% due to changes in work and production schedule.
As for industrial demand, \citet{DECC2012} and \citet{larsen2006effect} have cited energy audits, energy and environment management systems, and voluntary agreements as possible measures to reduce energy in energy intensive sectors, as it pushes the entire organization to follow an internationally recognized process that influences overall energy consumption. Furthermore, \citet{owen2010employee} declare a 12\% reduction potential through Community-Based Social Marketing, such as energy information and feedback systems. The findings of \citet{hansen_working4utah_2009} are partly transferred to industry, since office spaces are mostly utilizing the same lighting and equipment. 

The overall potential for the \textit{Low Ambition} and \textit{High Ambition} scenario used hereafter is summarized in Table \ref{tab:saving_potential_electricity}. It is derived by linking the reductions from literature with the respective shares of energy demand.\footnote{A more detailed description can be found in \citet{pjs_report}.} Again, a negative percentage refers to the reductions that are derived for the respective category and percentages in the column headings refer to the share of total conventional electricity demand, e.g. residential electricity demand accounts for approximately 25 percent of total conventional electricity demand. 

\begin{table}
\caption{Sufficiency potential for residential, commercial and industrial conventional electricity demand}
\label{tab:saving_potential_electricity}
\centering
\begin{tabular}{llcccc}
\midrule
\multicolumn{1}{c}{\textbf{Reduction potential}} &  \begin{tabular}[c]{@{}c@{}}\textbf{Residential}\\ \textbf{(25.3\%)}\end{tabular}  & \begin{tabular}[c]{@{}c@{}}\textbf{Commercial}\\ \textbf{(28.9\%)}\end{tabular} & \begin{tabular}[c]{@{}c@{}}\textbf{Industrial}\\ \textbf{(45.8\%)}\end{tabular} & \begin{tabular}[c]{@{}c@{}}\textbf{Total conventional}\\ \textbf{electricity}\end{tabular} \\\midrule
Low   Ambition          & -4.0\%                  & -5.5\%              & -7.0\%              & -5.8\%             \\
High Ambition         & -20.0\%                 & -23.4\%             & -17.5\%             & -19.9\%             \\ \midrule
\end{tabular}
\end{table}

\section{Methodology and Scenarios}\label{sec:methodology}
The derived potential for sufficiency measures is used as an input for an energy system model to investigate their impact on the supply side. Next, the methodical approach and the underlying data structure is briefly explained, followed by a detailed description of the modeled scenarios.

\subsection{Applied model}\label{subsec:model_data}
For the evaulation of sufficency measures, a linear bottom-up planning model is employed. The model optimizes the German energy system in a greenfield approach and is implemented in the open-source, Julia-based framework \textit{AnyMOD.jl}. The framework is developed to model complex energy systems with high spatial and temporal resolution. Main advantages of the framework include the flexibility to model energy carriers at different temporal resolutions \citep{goke2020anymod, goke2020anymodjl}.

Exogenous input to the model are cost assumptions, available technologies, conversion efficiencies, renewable potential and hourly time series for renewable availability and demand. To satisfy an exogenous final demand, the model decides on expansion and operation of technologies to generate, store, and transport energy carriers, while minimizing the sum of investment and operational costs. Since our focus in on a fully renewable energy system, the technology portfolio is limited to renewable technologies.

\begin{figure}[htbp]
    \centering
    \includegraphics[scale = 0.4]{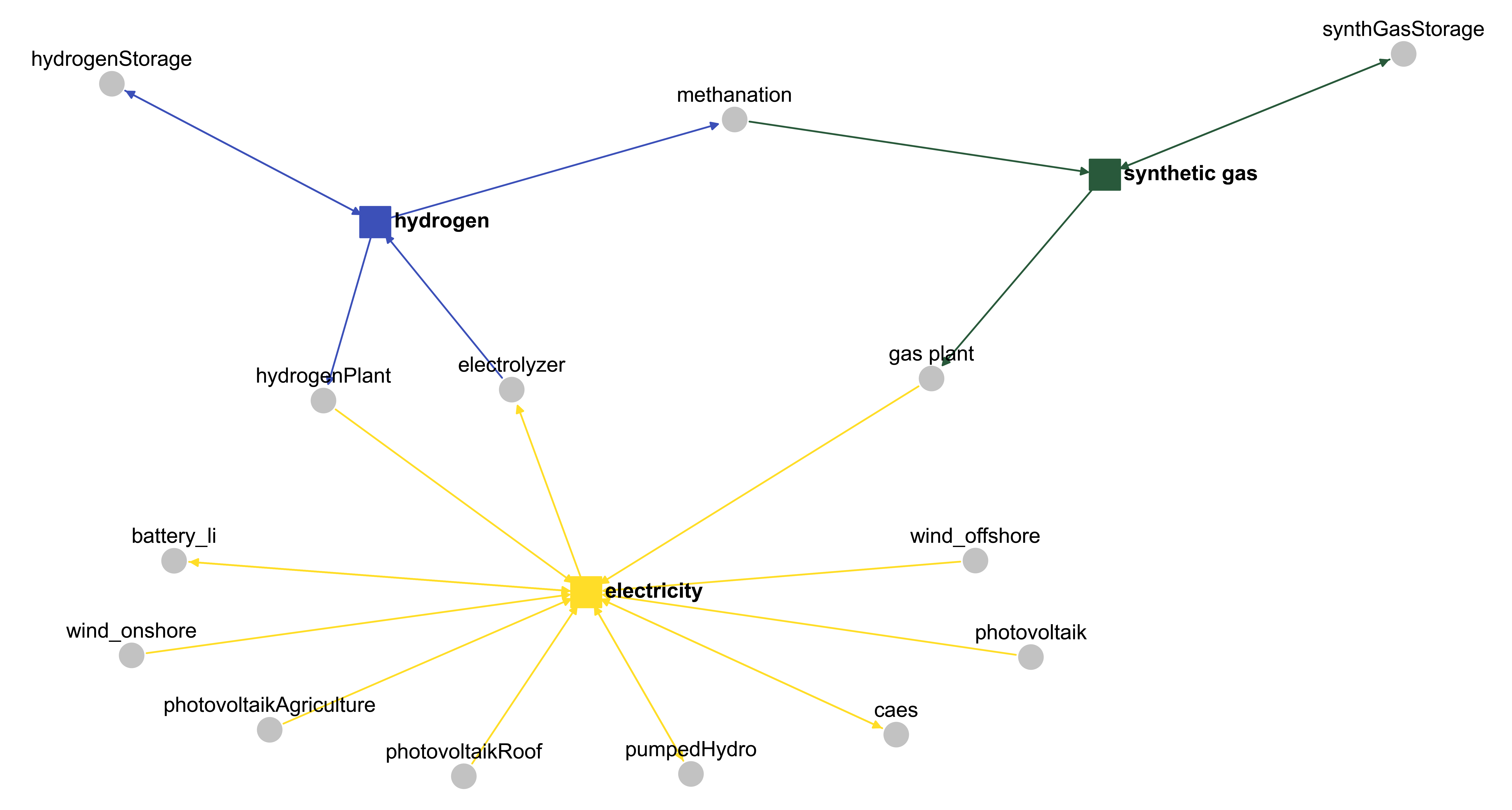}
    \caption{Overview of modeled energy carriers and technologies. }
    \label{fig:tech_nodes}
\end{figure}

Figure \ref{fig:tech_nodes} provides an overview of the considered energy carriers and technologies and how they are interlinked. In the figure, carriers are symbolized by colored squares and technologies by gray circles. Edges between technologies and carriers indicate their relation and entering edges of technologies refer to input carriers; outgoing edges refer to outputs. For example, a gas plant takes synthetic gas as an input to generate electricity. The modeled technologies include generation and storage technologies. Generation technologies supply a certain carrier, often by converting energy from one form to another (e.g. from kinetic to electrical or electrical to chemical energy). Available generation technologies are photovoltaic (openspace, rooftop and agricultural\footnote{Agricultural photovoltaic refers to solar panels that share space with conventional agricultural activity. The possible coexistence proves a high energetic potential on vast areas and dual-use opportunities.}), wind turbines (onshore and offshore), hydrogen and synthetic gas turbines, electrolyzers and plants for methanation. Available storage technologies are lithium-ion batteries, pumped hydro storages, gas storages and compressed air storages. Table \ref{app:generation_costs} and \ref{app:storage_costs} in the Appendix provide a comprehensive overview on the cost assumptions and data sources for generation and storage technologies. Figure \ref{fig:renewable_potential} in the Appendix shows the potential that limits the maximum capacity that can be installed for a specific technology \citep{auer2020development, pvagr}. 

\begin{figure}[htbp]
    \centering
    \includegraphics[scale=0.26]{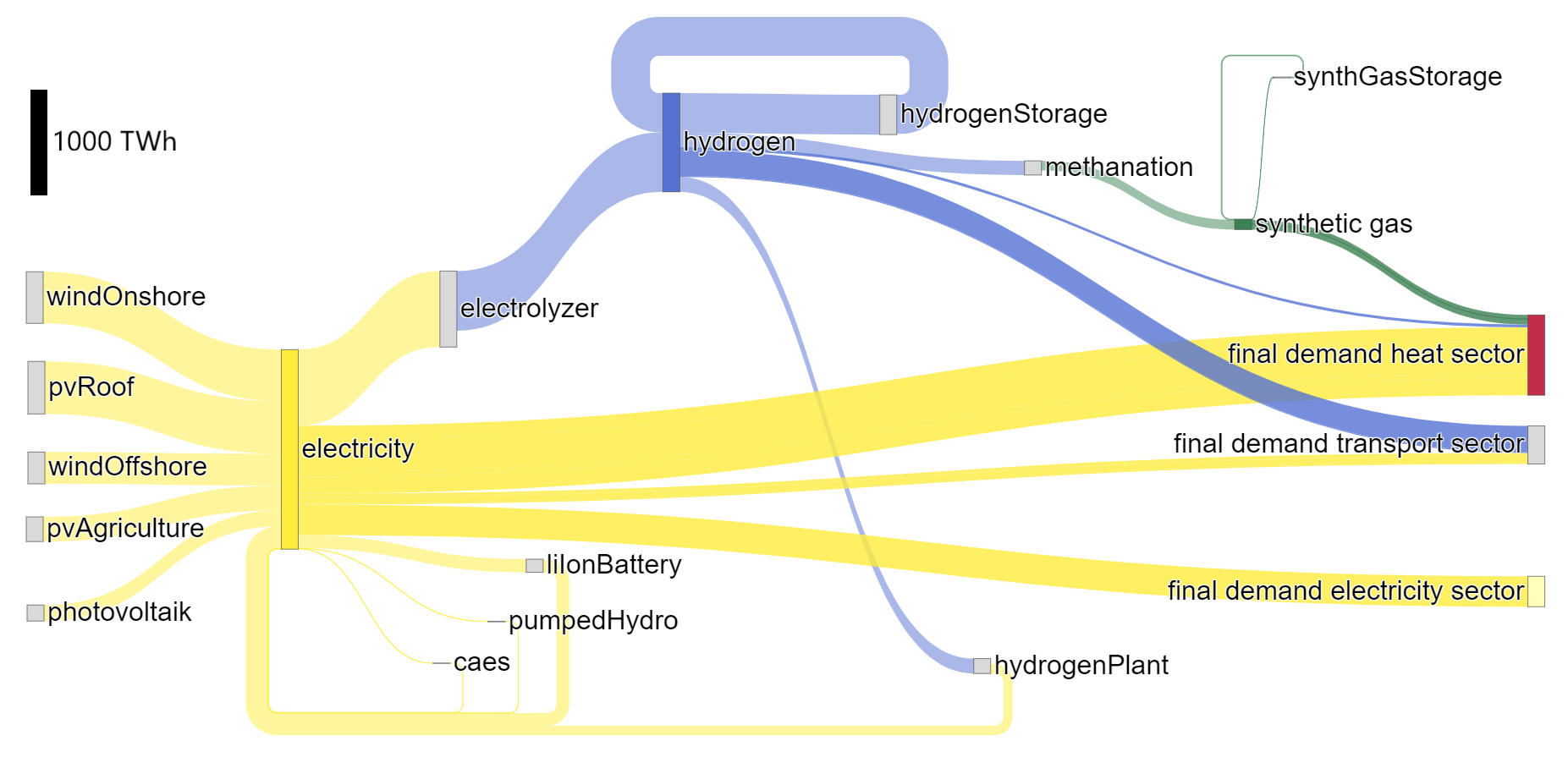}
    \caption{Energy flow diagram of the reference case.}
    \label{fig:base_sankey}
\end{figure}

To account for temporal variations in energy demand and renewable generation, both are implemented as hourly time series data from the openENTRANCE project \citep{auer2020development}.\footnote{For details see Figure \ref{fig:load_all_sectors} the Appendix.} Figure \ref{fig:base_sankey} shows a Sankey diagram visualizing the energy flow when solving the model in the reference case, meaning without any sufficiency measures. Final energy demand found on the right hand side of the diagram is separately provided for the heat, mobility, and electricity sector. The remainder of the diagram demonstrates how demand of these sectors is satisfied in the model by different technologies generating and storing energy.

Note that the stylized model does not consider import and exports of energy, which was adopted from other research and is a common approach when analyzing fundamental trade-offs in energy systems \citep{SCHILL2018156}. In addition, we assumed an unconstrained transport of energy within Germany. While these limitations reduce the applicability to real-world energy system, they focus the research on macro-level interactions between sufficiency-related demand reductions and the rest of the energy system. 

\subsection{Scenarios}\label{sec:scenarios}
\begin{figure}[htbp]
    \centering
    \includegraphics[scale=0.8]{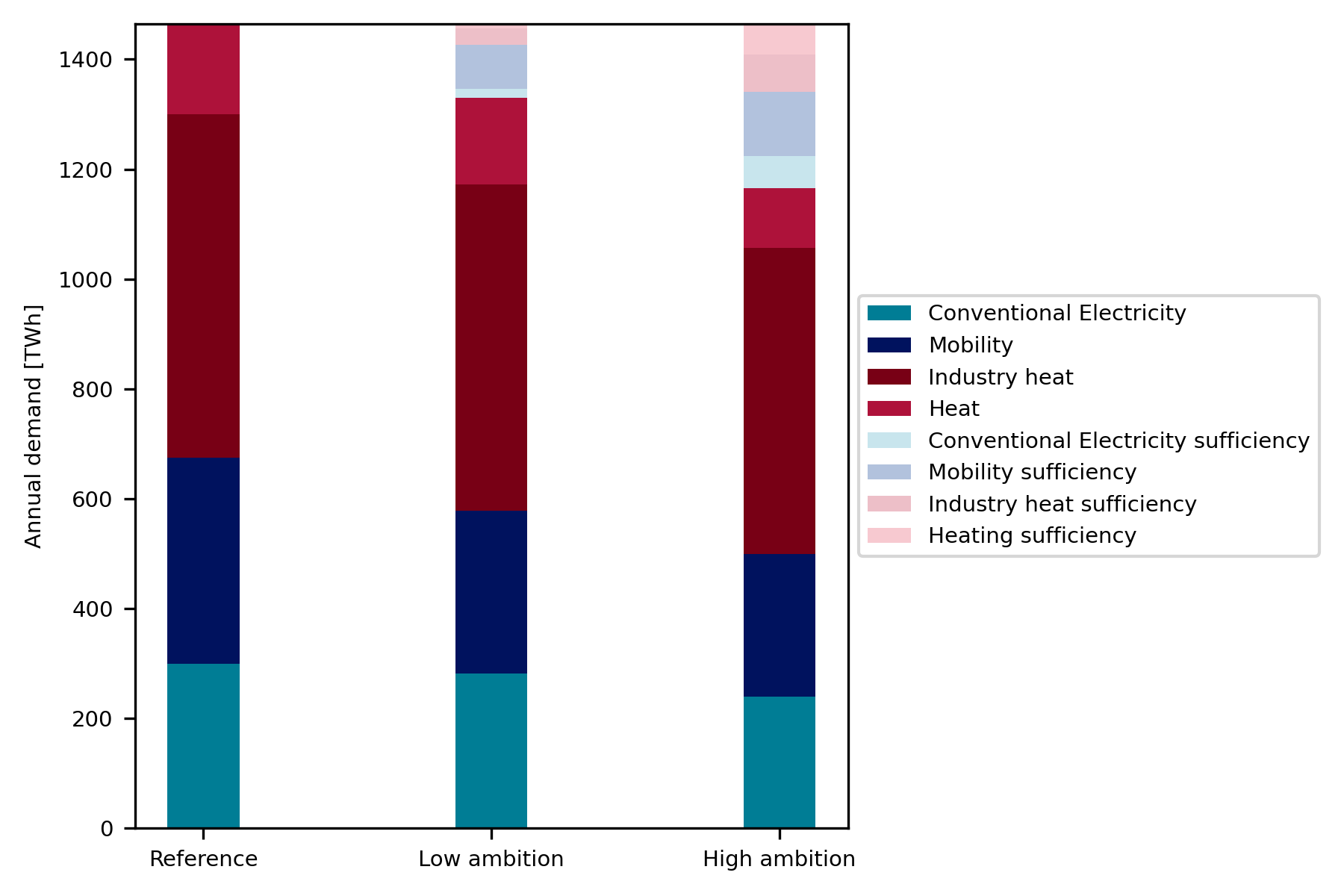}
    \caption{Visualization of the composition of the energy demand across scenarios}
    \label{fig:scnario_all_three__visu}
\end{figure}

To distinguish between different levels of behavioral change, the sufficiency measures are implemented in two scenarios. Figure \ref{fig:scnario_all_three__visu} presents the demand reductions assumed for the \textit{Low Ambition} and \textit{High Ambition} scenario with dark colors representing the energy demand that still needs to be met, and lighter colors representing demand reductions from sufficiency measures for each sector. The exact corresponding values are listed in Table \ref{tab:all_three_sector_low_high}. Percentages refer to the reduction relative to the individual sector demand, e.g. mobility reduction potential refers 30.9\% of the mobility energy demand. There are no specific assumptions on the temporal structure of demand reductions and relative reductions are applied equally to each hour of the year year, which translates to a proportional downward shift of the load curve. 

\begin{table}[htbp]
\caption{Sufficiency-based demand reductions in each sector.}
\label{tab:all_three_sector_low_high}
\centering
\begin{tabularx}{\linewidth}{lXXXXc}
\midrule
\textbf{Scenario} & \textbf{Conventional Electricity} &\textbf{Mobility}&\textbf{Residential/ Commercial Heat} & \textbf{Process Heat} & \textbf{Total}\\ \midrule
Low Ambition & -5.8 \%&-21.7\%\ & -5.0\%   & -4.9\%  & -9.4\%            \\ 
High Ambition & -19.9 \% & -30.9\% & -34.2\%  & -10.9\%  & -20.5\%        \\ \midrule
\end{tabularx}
\end{table}

\subsection{Sensitivity analysis}\label{peak_load_shedding} 
An additional sensitivity analysis investigates the impact of each sector individually.  In this case, only the assumptions of the \textit{High Ambition} scenario are applied for each sector individually to allow for a comparison of sector-specific effects on the supply side. As a result, the quantitative difference of sectoral reductions can be elaborated on.

\section{Results}\label{sec:results}
The results of the energy system model for the different scenarios can be used to estimate the impact of sufficency measures. Key results include cost savings and the impact on generation and storage capacities.

\subsection{Sufficiency scenarios}
Demand reductions from the \textit{Low Ambition} and \textit{High Ambition} scenario result in a reduction of system costs by 11.3\% and 25.6\% respectively, as listed in Table \ref{tab:combined_total_cost_table}. There is an overproportional decrease in costs from \textit{Low Ambition} to \textit{High Ambition}, because demand reduces by 11.1\% in between the scenarios, but cost by 14.3\%. This is due to the higher reduction of the demand for synthetic gas in the \textit{High Ambition} scenario , resulting in larger cost savings given the loss in inefficiencies from electrolysis and methanation versus direct use of electricity. 

\begin{table}
\centering
\caption{Total annualized system costs and demand reductions in the reference, \textit{Low Ambition} and \textit{High Ambition} scenario.}
\label{tab:combined_total_cost_table}
\begin{tabular}{lccc}
\toprule
& \textbf{Reference} & \textbf{Low}  & \textbf{High}  \\ \midrule
Total costs {[}M€{]} & 119,399 & 105,897 & 88,867 \\
Cost  & 0\%  & -11.3\%  & -25.6\%    \\
Total demand & 0\%   & -9.4\%   &  -20.5\%  \\  \bottomrule
\end{tabular}
\end{table}

\begin{figure}
\begin{subfigure}{.5\textwidth}
  \centering
  \includegraphics[scale=0.9]{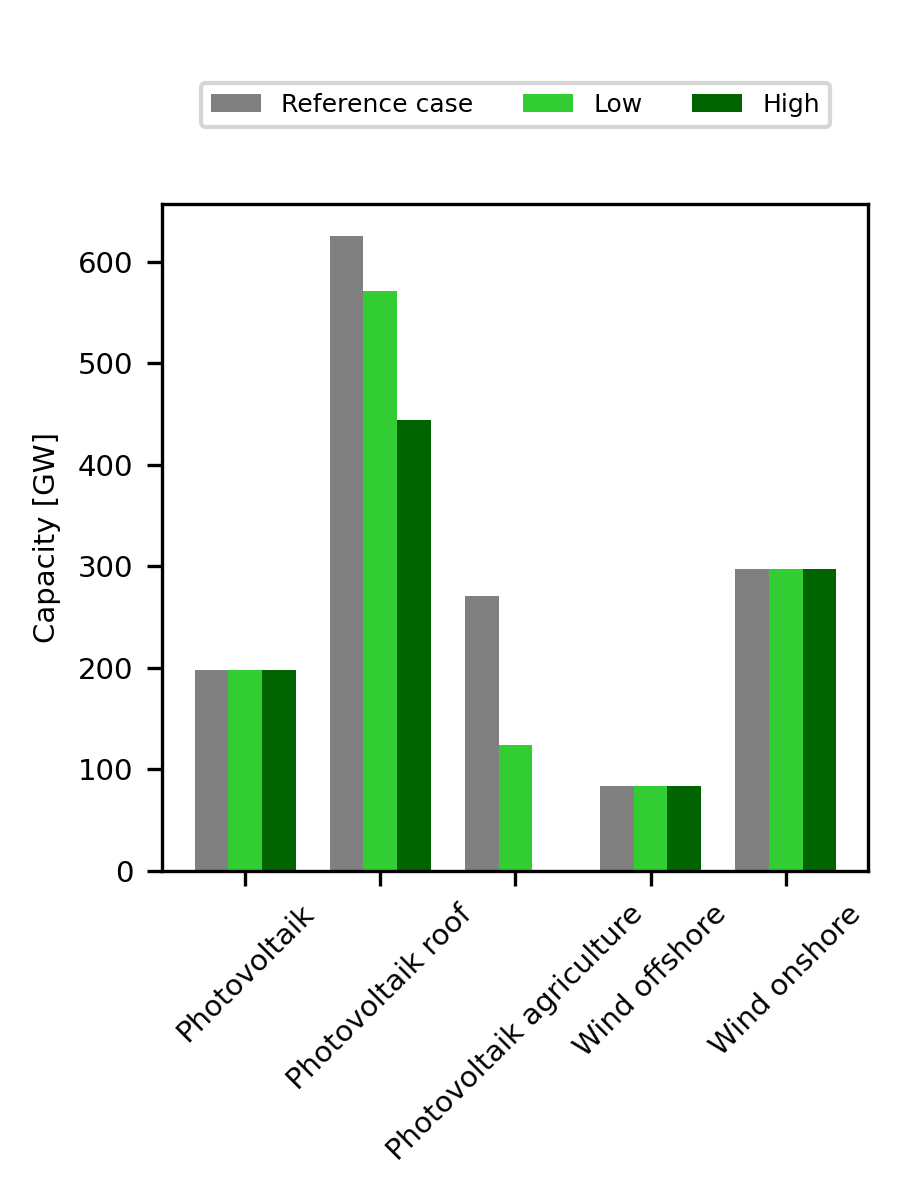}
  \caption{Capacity of renewable technologies}
  \label{fig:8.4_combined_renewable_summary_scenarios}
\end{subfigure}%
\begin{subfigure}{.5\textwidth}
  \centering
  \includegraphics[scale=0.9]{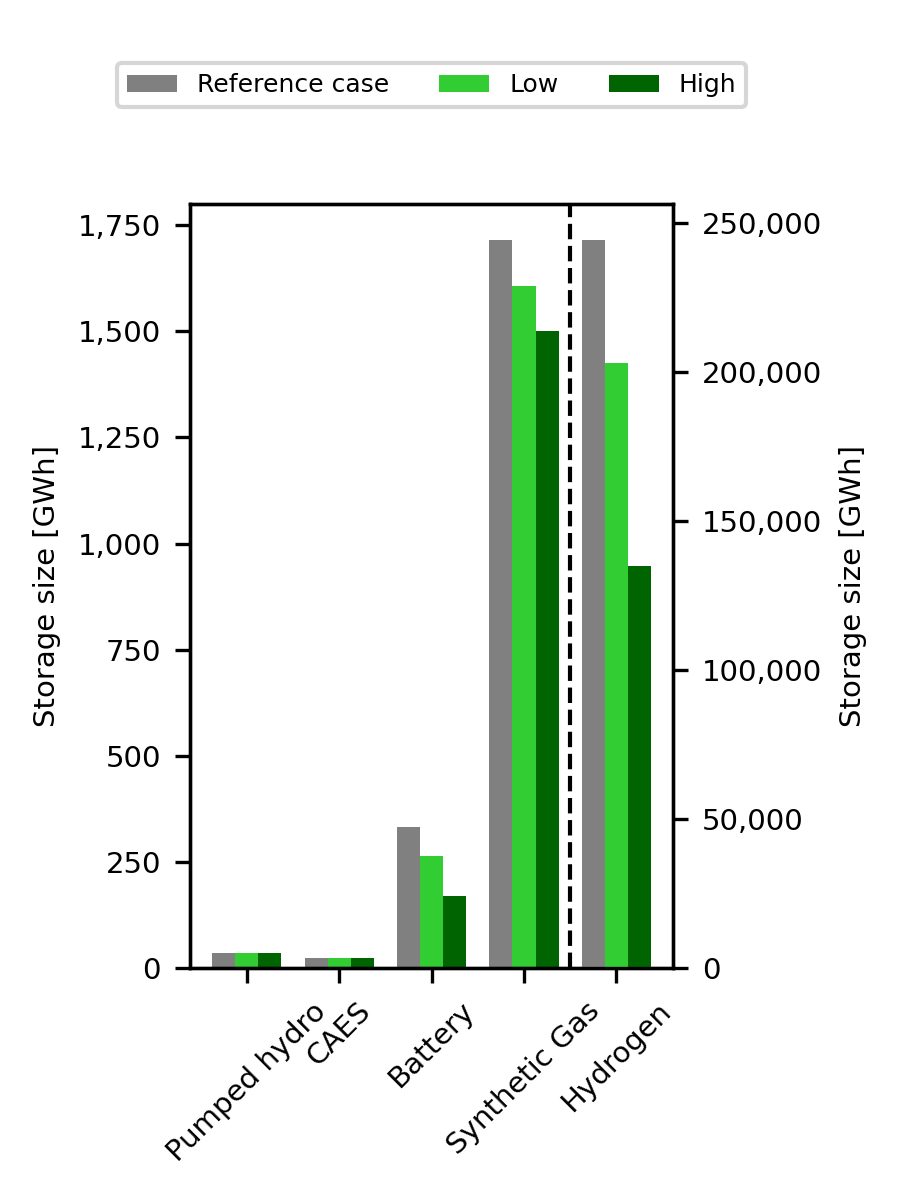}
  \caption{Size of storage technologies}
  \label{fig:8.4_combined_stor_summary_scenarios}
\end{subfigure}
\caption{Model results of scenarios. Installed renewable generation capacities (a) and storage size (b) for \textit{Low Ambition} and \textit{High Ambition}.}
\label{fig:combined_results_capa}
\end{figure}

The demand reductions in the \textit{Low Ambition} and \textit{High Ambition} scenario lead to an overall renewable capacity reduction of 13.6\% and 30.6\% respectively, as well as a storage size reduction of 16.8\% and 44.5\%. Figure \ref{fig:combined_results_capa} details the capacity reductions in each technology, reporting that the need for capacity and storage is decreasing relative to demand. The most significant storage reduction comes from hydrogen storage, as shown in Figure \ref{fig:8.4_combined_stor_summary_scenarios}, because the need for long term storage decreases. Necessary generation capacity is reduced for agricultural and rooftop photovoltaic, probably explained by high capacity costs due to the limited availability of solar radiation. In the \textit{High Ambition} scenario, the overall photovoltaic capacity is reduced by 41.3\%, and agricultural photovoltaic is not necessary at all. 

\subsection{Sensitivity analysis}
\begin{figure}[htbp]
	\centering
	\includegraphics[scale=0.8]{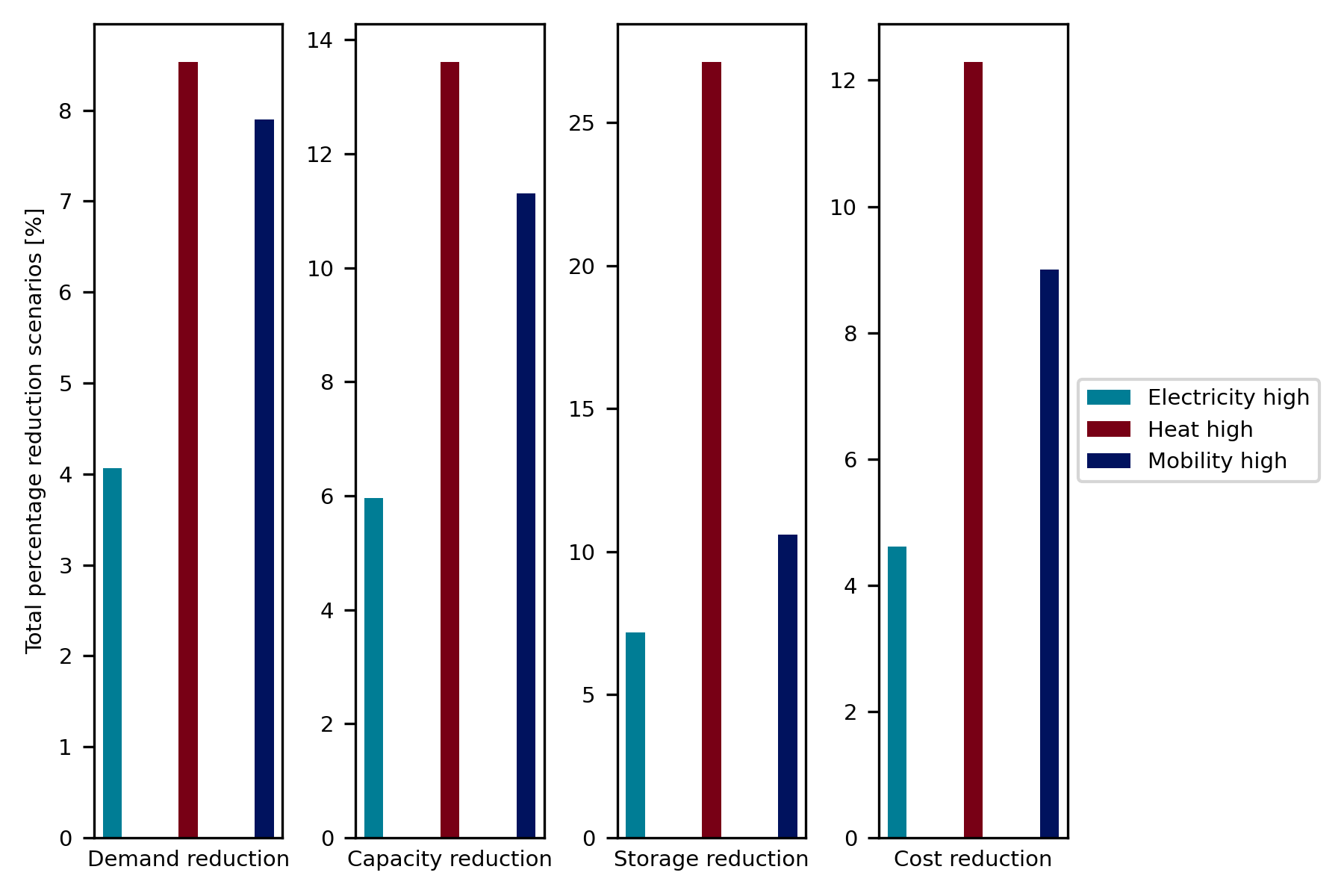}
	\caption{Impact of assumptions in the high ambitions scenario on each sector}
	\label{fig:seperate_sector_diff}
\end{figure}

Figure \ref{fig:seperate_sector_diff} compares effects of the individual sectors on generation capacities, storage sizes and total annualized system costs. In each case, only one sector's demand is reduced, while the other two remain unchanged. The heat and transport sector have a demand reduction potential of about 8\% of the total energy demand, visible in the left subplot of Figure \ref{fig:seperate_sector_diff}. It is noticeable that reductions of the same magnitude lead to storage reduction of approximately 11\% for the transport sector and 27\% for the heat sector. Therefore, substantial cost savings for storage technologies can be archieved, in particular by reducing the required capacities of li-ion batteries, saving 1,500 M€ of annualized costs.

Because of high peak loads, sufficiency measures in the heating sector show the highest impact on the systems supply side, as demand reductions lead to overproportional cost reductions. For instance, residential heat indicates cost savings 1.6 times larger than the applied demand reduction, and process heat shows cost savings 1.4 times larger than the applied demand reduction. In comparison, sufficiency measures in the transport and electricity sector lead to cost savings that are only 1.1 times larger than the demand reduction. In total, the heating sector reaches the highest cost reduction of 12.3\% compared to the reference case.

\begin{table}[htbp]
	\centering
	\caption{Overview on total system cost and demand reductions by sector}
	\label{tab:cost_demand_ratio}
	\begin{tabular}{lcccc}
		\toprule
		\textbf{Sector} &\textbf{Total System Costs {[}M€{]}} & \textbf{Cost reduction} &\textbf{Demand reduction} \\ \midrule
		Reference & 119,399 & 0.0\%& 0.0 \%\\\midrule
		Electricity & 113,888&4.6\%& 4.1\%\\
		Transport & 108,656 & 9.0\% & 8.0\% \\ 
		Heat (a+b) & 104,731 & 12.3\% & 8.3\%  \\\midrule
		a. Heat residential\protect\footnotemark & 111,889 & 6.3\%& 3.9\%  \\ 
		b. Process heat & 112,229 & 6.0\%&4.5\% \\ \bottomrule
	\end{tabular}
\end{table}

\footnotetext{"Heat residential" includes the private and commercial heat demand at this point.}

\section{Conclusion}\label{sec:conclusion}
The decarbonization necessitates a rapid transformation of energy systems, which includes a massive expansion of renewable generation capacity and the utilization of all available flexibility options, such as storages, grid expansion and sector coupling. Therefore, large-scale energy system models that include all sectors are increasingly employed to understand the complex relationships of a 100\% renewable energy system. While many studies model a decarbonized energy system, few include changes on the demand side in their scenarios. This research has therefore explicitly investigated the potential of behavioral changes to reduce final energy demand, referred to as sufficiency measures.

Based on literature, the potential for sufficiency measures was quantified for two different levels of behavioral change in terms of ambition. These potentials where used as the scenarios \textit{Low Ambition} and \textit{High Ambition}. An overall sufficiency potential of 9.4\% to 20.5\% has been identified respectively. Reference energy demand of 1465 TWh is thus reduced by 138 TWh to 1327 TWh in \textit{Low Ambition} and by 300 TWh to 1165 TWh in \textit{High Ambition}. The heat and transport sector show sufficiency potentials of 124 to 115 TWh respectively, followed by the electricity sector with up to 60 TWh reduction potential. The impact these reductions have on the supply side of the energy system was evaluated using a cost-minimizing bottom-up planning model for the scenarios \textit{Low Ambition} and \textit{High Ambition}. Results when taking sufficiency measures were compared against a reference case without any behavioral changes. The results show that sufficiency measures can achieve cost savings of 11.3\% to 25.6\% and reduce generation and storage capacity by 30.6\% and 44.5\% in \textit{Low Ambition} and \textit{High Ambition}. The potential cost savings from sufficiency are biggest in the heat sector due to its high peak loads. Final energy demand in the heat sector can be reduced by 8.3\%, resulting in overproportional cost savings of 12.3\%. In comparison, demand reduction of 4.1\% in the electricity and 8\% in the transport sector show only a cost reduction of 4.6\% and 9\% respectively.

Our assessment of sufficiency measures is still subject to a number of limitations influencing the obtained results. First of all, the applied energy system model is stylized. For example, it still neglects cross-border exchange of energy and is limited to a greenfield approach. Additionally, in many cases literature only provides rough estimates for the potential of certain sufficiency measures and transferring these estimates into reductions of final energy demands adds to the imprecision. Furthermore, the identified demand reductions were implemented without assessing the political feasibility or the impacts on socio-economic indicators, such as job creation and distributional effects. Some potentials seem more difficult to realize, e.g. a lower room temperature. Other potentials seem already realistic today, as Covid-19 has shown that realizing high shares of home office and telemeetings are possible.

In conclusion, our quantitative evaluation suggests significant benefits from considering sufficiency measures for decarbonization. On the one hand, this means future research should dedicate further efforts to quantify the sector-specific potential for sufficiency measures. Results in this study were based on a literature review and should be considered first estimates that can be further refined by sector-specific analysis. For example, it appears promising to combine detailed models of residential heat demand with a comprehensive energy system model for a highly resolved analysis of sufficiency measures in this area. On the other hand, scenarios for public policy should take greater account of sufficiency measures and future research should evaluate adequate policy instruments to promote sufficiency. In policy, a framework to promote and increase the likelihood of sufficient behavior is still missing. For example, policies that target sufficient behavior could for instance aim at the size of end-use equipment, discourage individual property in favor of a shared economy and promote shared living space or reduce aviation frequency and length \citep{SAMADI2017126}. Consequently, sufficiency could be included in at least one scenario in every scenario-based analysis to depict possible decarbonization pathways' bandwidth accurately in scientific research.


\section*{Declaration of competing interest}
The authors declare that they have no known competing
financial interests or personal relationships that could have
appeared to influence the work reported in this paper.

\section*{Acknowledgements}
We thank Jens Weibezahn, Leonard Göke, Mario Kendziorski and Christian von Hirschhausen for initiating this research; special thanks go to Carl-Christian Klötzsch, Eric Rockstädt, Michel Kevin Caibigan, Morteza Lavaiyan and Nithish Kini Ullal for their collaboration in this project. The usual disclaimer applies.

This research did not receive any specific grant from funding agencies in the public, commercial, or
not-for-profit sectors. 

\appendix
\renewcommand{\thetable}{\Alph{section}.\arabic{table}}
\setcounter{table}{0}
\renewcommand{\thefigure}{\Alph{section}.\arabic{figure}}
\setcounter{figure}{0}

\section{Sufficiency-based demand reductions}

\begin{table}[htbp]
\caption{Sufficiency-based demand reductions from literature in the sector conventional electricity. Measures refer to the causes that induce demand reductions, the value range refers to the reductions range that was identified in literature. Measures can either be politically promoted or occur due to changes in socio-economic norms.}
\label{tab:electricity}
\centering

\begin{tabularx}{\linewidth}{lllX}
\toprule
  \textbf{Sub-category}  & \textbf{Measures} & \textbf{Value range} & \textbf{Source} \\ \midrule
 \multicolumn{4}{l}{\textit{Residential}}     \\ \midrule
 
   & Direct feedback & -5.0\% to -15.0\% & \citet{martiskainen_affecting_2007} \\ \cline{2-4} 
   & \begin{tabular}[c]{@{}l@{}}Direct feedback applications \\ (Europe and N.America)\end{tabular} & -9.0\% & \citet{en12193788} \\ \cline{2-4} 
   & \begin{tabular}[c]{@{}l@{}}Indirect feedback applications \\ (Europe and N.America)\end{tabular} & -4.0\% & \citet{en12193788} \\ \cline{2-4} 
   & Goal setting & -4.5\% & \citet{martiskainen_affecting_2007} \\ \cline{2-4} 
   & Goal setting with feedback & -15.1\% & \citet{martiskainen_affecting_2007} \\ \cline{2-4} 
   & \begin{tabular}[c]{@{}l@{}}Feedback through smart meters \\ and time-use tariffs (Ireland)\end{tabular} & -1.8\% & \citet{CARROLL2014234} \\ \cline{2-4} 
  & Feedback (peak reduction) & -7.8\% & \citet{CARROLL2014234} \\ \cline{2-4} 
   \multirow{-9}{*}{\begin{tabular}[c]{@{}l@{}}Behavioral change \\ through intervention\end{tabular}} & Change of human behavior & -20.0\% & \citet{burger2009identifikation} \\ \cline{1-4}
   & Sufficiency  in lighting and appliances & -15.0\% to -20.0\% & \citet{umweltbundesamt_konzept_2015} \\ \cline{2-4} 
  \multirow{-2}{*}{\begin{tabular}[c]{@{}l@{}}Change in user behavior \\ through   intrinsic motivation\end{tabular}} & Change of human behavior & -20.0\% & \citet{burger2009identifikation} \\ \midrule

 \multicolumn{4}{l}{\textit{Commercial}}     \\ \midrule

   & Group level feedback & -7.0\% & \citet{CARRICO20111} \\ \cline{2-4} 
   & Goal setting & -12.9\% & \citet{nilsson2015energy} \\ \cline{2-4} 
  \multirow{-3}{*}{\begin{tabular}[c]{@{}l@{}}Behavioral change \\ through intervention\end{tabular}} & Goal   setting with feedback etc. & -5.5\% and -6.0\% & \citet{nilsson2015energy} \\ \cline{1-4} 
  \begin{tabular}[c]{@{}l@{}}Behavioral change \\ through use of technology\end{tabular} & \begin{tabular}[c]{@{}l@{}}Energy   information system + \\ social marketing - feedback \\ (Community Based Social   Marketing)\end{tabular} & -12.0\% & \citet{owen2010employee} \\ \cline{1-4}
  \begin{tabular}[c]{@{}l@{}}Revolutionary changes \\ through legislations\end{tabular} & \begin{tabular}[c]{@{}l@{}}Four-day week/shorter working \\ time/less production\end{tabular} & -10.5\% & \citet{hansen_working4utah_2009} \\ \midrule

 \multicolumn{4}{l}{\textit{Industrial}}     \\ \midrule

   & \begin{tabular}[c]{@{}l@{}}Energy Audits \\ (Denmark)\end{tabular} & -7.0\% to -20.0\% & \citet{larsen2006effect} \\ \cline{2-4} 
  \multirow{-4}{*}{\begin{tabular}[c]{@{}l@{}}Behavioral change \\ through intervention\end{tabular}} & \begin{tabular}[c]{@{}l@{}}Energy information system + \\ social marketing - feedback \\ (Community Based Social Marketing)\end{tabular} & -12.0\% & \citet{owen2010employee} \\ \cline{1-4}
  \begin{tabular}[c]{@{}l@{}}Revolutionary changes\\ through legislations\end{tabular} & \begin{tabular}[c]{@{}l@{}}Four-day week/ shorter working time/\\ less production\end{tabular} & -10.5\% & \citet{hansen_working4utah_2009} \\ \bottomrule
\end{tabularx}%

\end{table}

\clearpage

\begin{table}[ht]
\caption{Sufficiency-based demand reductions from literature in the sector mobility. Measures refer to the causes that induce demand reductions, the value range refers to the reductions range that was identified in literature. Measures can either be politically promoted or occur due to changes in socio-economic norms.}

\begin{tabularx}{\linewidth}{Xll}
\midrule
\textbf{Measure}  & \textbf{Value range} & \textbf{Source}          \\ \midrule
\multicolumn{3}{l}{\textit{Bicycles}}     \\ \midrule
Modal shift from passenger car
to cycling             & -3.5\% to  -10.0\%       & \makecell[l]{\citet{umweltbundesamt_konzept_2015}\\ \citet{van2018potential}}  \\
Increased level of investment in bike   infrastructure       & +50.0\% to +100.0\%        & \citet{VENTURINIb}\\
Increased share of E-Bikes in total         & +1.0\% to +50.0\%          & \citet{VENTURINIb}   \\  \midrule
\multicolumn{3}{l}{\textit{Passenger cars}}         \\  \midrule
Replacing business trips with telemeetings  & -40.0\% to -60.0\%       & \citet{umweltbundesamt_konzept_2015}                   \\
Smaller passenger cars through regulation   & -7.5\%               & \citet{umweltbundesamt_konzept_2015}                \\
Reduction of motorized individual transportation                      & -30.0\%                & \citet{Fraunhofer2020}    \\
Reduced commuting demand through teleworking& -1.0\% to -20.0\%        &
\makecell[l]{\citet{van2018potential}\\ \citet{VENTURINIb}} \\
Increased load factor for every commute car trip (carpooling)         & Load factor 2     & \makecell[l]{\citet{van2018potential}\\ \citet{VENTURINIb}} \\  \midrule
\multicolumn{3}{l}{\textit{Public transport}}                       \\ \midrule
Modal shift to public transport for all commuting demand              & -100.0\%               & \citet{van2018potential} \\
Reduced traveling time of public transport  & -1.0\% to -10.0\%        & \citet{VENTURINIb}   \\ \midrule
\multicolumn{3}{l}{\textit{Aviation}}  \\ \midrule
Reduction of aviation & -55.0\%   & \citet{Fraunhofer2020} \\
Reduction of private aviation               & -50.0\%                & \citet{umweltbundesamt_konzept_2015}               \\
Avoid flights that can be replaced by another transport mode \textless 10h     & -25.0\%                & \citet{van2018potential} \\
Replace intercontinental leisure flights with intra-EU trips & -50.0\% & \citet{van2018potential}  \\ \midrule
\end{tabularx}
\label{tab:mobility}
\end{table}

\begin{table}[htbp]
\caption{Sufficiency-based demand reductions from literature in the sector heat. Measures refer to the causes that induce demand reductions, the value range refers to the reductions range that was identified in literature. Measures can either be politically promoted or occur due to changes in socio-economic norms.}
\label{tab:entire_heat_literature}

\begin{tabularx}{\linewidth}{Xll}
\midrule
\textbf{Measure}& \textbf{Value range} & \textbf{Source}  \\ \midrule
\multicolumn{3}{l}{\textit{Space heating demand}}\\ \midrule
Lowering average room temperature by 1-2\textdegree C &  -4.4\% to -9.0\%   & 
\makecell[l]{\citet{umweltbundesamt_konzept_2015}\\ \citet{marshall_combining_2016}} \\
Turning down thermostat by 1\textdegree C & -13.0\% & \citet{palmer2012much} \\
Decreasing living space per person & -24.9\% to 35.7\%  & \citet{BierwirthThomas2019}  \\ \midrule
\multicolumn{3}{l}{\textit{Hot water consumption}} \\ \midrule
Water efficient shower heads & -50.0\%  & \citet{palmer2012much}                   \\
Feedback system about showering time & -5.0\% to -10.0\% & \citet{toulouse_2018_products}\\
Shorter and less frequent showering   & -20.0\% to -30.0\%  & \citet{palmer2012much} \\ 
Adjusting water consumption & -70.0\% & \citet{lehmann2015stromeinspareffekte}\\  \midrule
\multicolumn{3}{l}{\textit{Process heat low temperature}} \\ \midrule
Decreasing food waste & -8.6\% to -13.2\% & \makecell[l]{\citet{lebensmittel2019}\\ \citet{ vita2019environmental}} \\ \midrule
\multicolumn{3}{l}{\textit{Process heat mid temperature}}    \\ \midrule
Increasing plastic recycling & -1.4\% to -2.1\%  &
\makecell[l]{\citet{negawatt}\\ \citet{ uba_kunststoffe}\\ \citet{chemiewirtschaft}} \\
\makecell[l]{Extending useful life of products and\\ establishing service-based sharing economy}& -3.0\% to -8.2\% & \citet{obsolescenceUBA, vita2019environmental} \\
\makecell[l]{Modal shift construction products and \\ reduced construction materials}& -0.7\% to -1.7\% & \citet{hertwich2019material} \\ \midrule
\multicolumn{3}{l}{\textit{Process heat high temperature}}    \\\midrule
\makecell[l]{Modal shift construction products and \\ reduced construction materials}&  -3.0\% to -7.6\% & \citet{hertwich2019material}     \\
\midrule
\end{tabularx}
\end{table}

\setcounter{table}{0}
\setcounter{figure}{0}

\section{Model input data}

\begin{figure}[htbp]
    \centering
    \includegraphics[scale = 0.7]{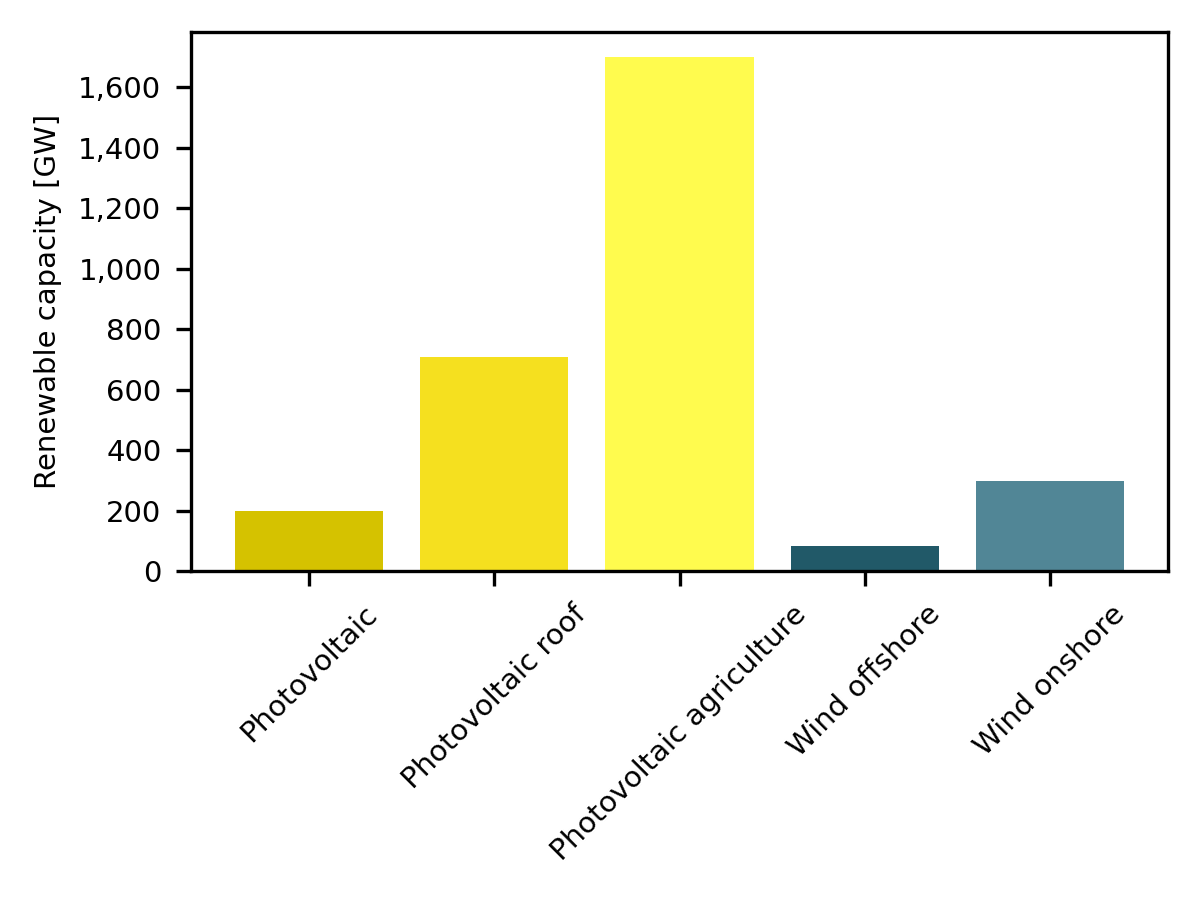}
    \caption{Renewable potentials in Gigawatts by technology. Potentials are mainly restricted by land availability. The model can invest into each technology until the limits are met. Own illustration based on \citet{auer2020development}}
    \label{fig:renewable_potential}
 \end{figure}

\begin{figure}[htbp]
    \includegraphics[scale=0.6]{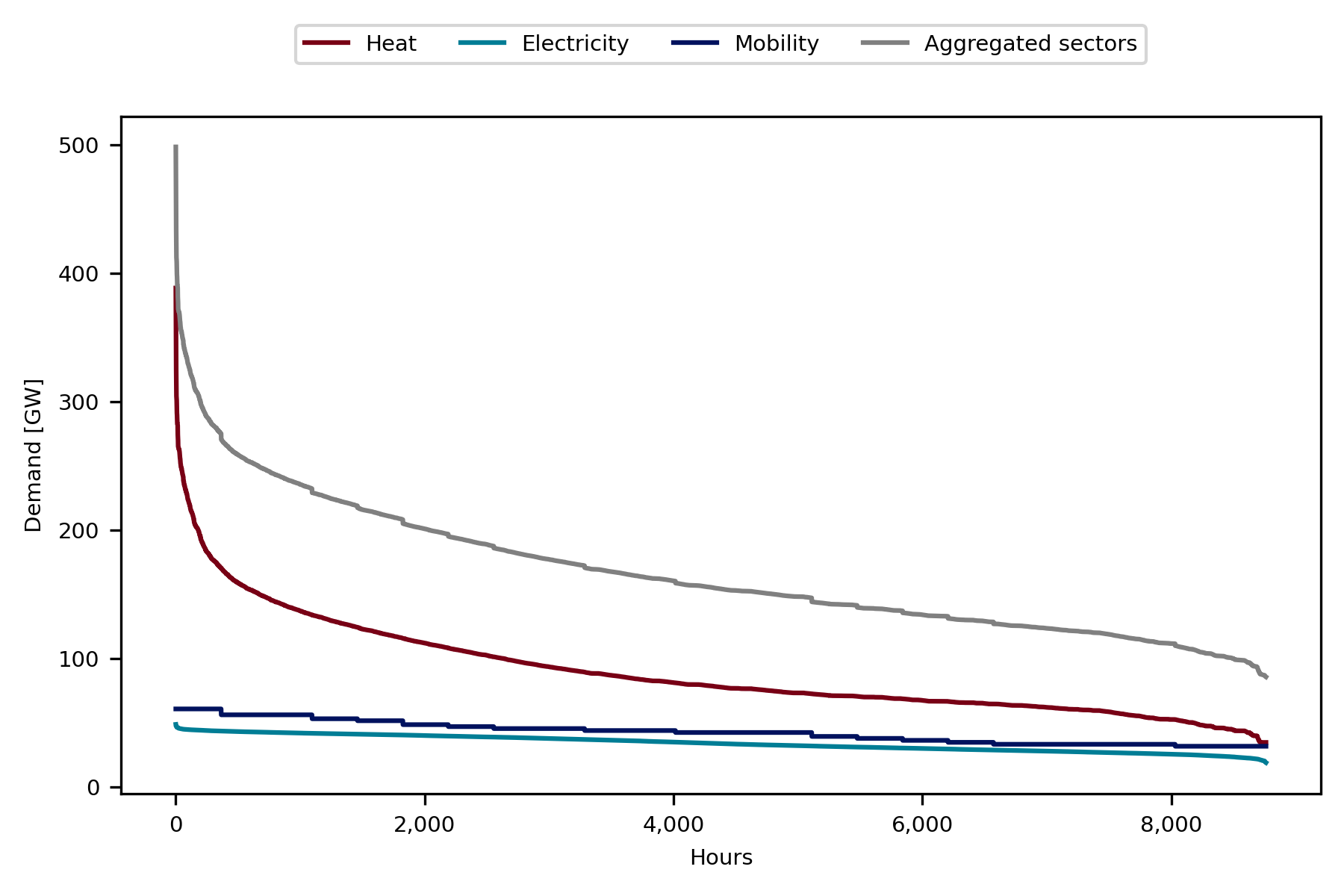}
    \centering
    \caption{Sectoral and aggregated load curve in reference case. Hourly load in Gigawatts [GWh] is depicted in descending order. Own illustration based on \citet{auer2020development}} 
    \label{fig:load_all_sectors}
\end{figure}{}

\begin{table}[htbp]
\caption{2035 Technology cost assumptions}
\label{app:generation_costs}
\begin{tabularx}{\linewidth}{lXXXl}
\toprule
\textbf{Technology} & \textbf{Investment Costs {[}€/kW{]}} & \textbf{Operating Costs {[}€/kW{]}} & \textbf{Lifetime {[}Years{]}} & \textbf{Source} \\ \midrule
ccgtGas        & 345  & 8.6  & 30 & \citet{auer2020development} \\
ccgtHydrogen   & 185  & 3.3  & 30 & \citet{auer2020development} \\
Methantion     & 865  & 18   & 30 & \citet{osmose_tr}          \\
Electrolyzer   & 543  & 14.6 & 30 & \citet{osmose_tr}          \\
PV             & 407  & 7.9  & 25 & \citet{auer2020development} \\
PV Rooftop     & 594  & 11.5 & 25 & \citet{auer2020development} \\
PV Agriculture & 814  & 7.9  & 25 & \makecell[X]{\citet{renewable_cost}\\ \citet{pvagr}} \\
Wind Onshore   & 1200 & 30   & 25 & \citet{auer2020development} \\
Wind Offshore  & 3111 & 100  & 25 & \citet{auer2020development} \\ \bottomrule
\end{tabularx}
\end{table}

\begin{table}[htbp]
\caption{2035 Storage cost assumptions}
\label{app:storage_costs}
\begin{tabularx}{\linewidth}{lXXXl}
\toprule
\textbf{Technology} & \textbf{Investment Costs {[}€/kWh{]}} & \textbf{Investment Costs {[}€/kW{]}} & \textbf{Lifetime {[}Years{]}} & \textbf{Source} \\ \midrule
Li-Ion Battery & 218  & 84.2 & 18 & \citet{osmose_tr}                              \\
Pumped Hydro   & 10   & 745  & 60 & \citet{osmose_tr}                              \\
Gas Storage    & 0.1  & 0.1  & 30 & \citet{osmose_tr}                               \\
CAES           & 26.4 & 455  & 30 & \citet{osmose_tr}                              \\ \bottomrule
\end{tabularx}
\end{table}

\bibliographystyle{unsrtnat}
\bibliography{literature}
\end{document}